\documentclass{ws-ijmpa}
\usepackage[super,compress]{cite}
\begin{document}

\def \be {\begin{equation}}
\def \ee {\end{equation}}
\def \ba {\begin{array}}
\def \ea {\end{array}}
\def \bea{\begin{eqnarray}}
\def \eea{\end{eqnarray}}

\def \a {\alpha}
\def \b {\beta}
\def \g {\gamma}
\def \G {\Gamma}
\def \d {\delta}
\def \D {\Delta}
\def \e {\epsilon}
\def \ve {\varepsilon}
\def \m {\mu}
\def \n {\nu}
\def \k {\kappa}
\def \l {\lambda}
\def \L {\Lambda}
\def \s {\sigma}
\def \S {\Sigma}
\def \r {\rho}
\def \o {\omega}
\def \O {\Omega}
\def \th {\theta}
\def \Th {\Theta}
\def \t {\tau}
\def \z {\zeta}

\def \p {\partial}
\def \f {\frac}
\def \na {\nabla}
\def \da {\Box}
\def \nn {\nonumber}
\def \ol {\overline}
\def \scl {\ell}
\def \ma {\mathcal}
\def \mb {\mathbb}
\def \mc {\mathcal}
\def \wh {\widehat}

\def \lt {\left}
\def \rt {\right}
\def \sl {\slashed}
\def \la {\leftarrow}
\def \ra {\rightarrow}
\def \lra {\leftrightarrow}
\def \sr {\sqrt}
\def \td {\tilde}
\def \wt {\widetilde}
\def \hs {\hspace}
\def \pp {\propto}
\def \inf {\infty}
\def \dd {\mathrm{d}}

\title{Harrison metrics for the Schwarzschild black hole}

\author{Fang-Fang Yuan}

\address{Institute of Theoretical Physics, Beijing University of Technology,
Beijing 100124, China\\ffyuan@emails.bjut.edu.cn}

\author{Yong-Chang Huang}

\address{Institute of Theoretical Physics, Beijing University of Technology,
Beijing 100124, China\\ychuang@bjut.edu.cn}

\maketitle

\begin{abstract}
Based on the hidden conformal symmetry, some authors have proposed a Harrison metric for the Schwarzschild black hole. We give a procedure which can generate a family of Harrison metrics starting from a general set of $SL(2, \mathbf R)$ vector fields. By analogy with the subtracted geometry of the Kerr black hole, we find a new Harrison metric for the Schwarzschild case. Its conformal generators are also investigated using the Killing equations in the near-horizon limit.
\end{abstract}

\section{Introduction}

The Kerr/CFT correspondence \cite{Guica0809} has attracted a lot of attention in recent years \cite{Compere1203}. Compared to the extremal cases, the non-extremal black holes are still difficult to tackle with using the conformal field theory or related techniques. However, a hidden conformal symmetry has been discovered in the near-region, low-frequency limit of the Kerr black hole \cite{Castro1004}. It is related to the scalar wave equation rather than the symmetry structure of spacetime geometry. This powerful approach has been applied to many different black holes.

The hidden conformal symmetry of the Schwarzschild black hole was investigated in Ref. \citen{Bertini1106}. In contrast with most black holes, only one set of the $SL(2,\mathbf R)$ generators could be defined. By applying an $SU (2,1)$ transformation, i.e., a Harrison or Kinnersley transformation \cite{{Harrison1968},{Kinnersley1973}}, on the Schwarzschild black hole, a new metric was found which has the same structure as the Bertotti-Robinson spacetime. Its Killing vectors were shown to reproduce the hidden conformal symmetry generators. The quasinormal modes were also studied using the operator method in Ref. \citen{Chen1009}. For subsequent developments, see Refs. \citen{{Franzin1107},{Li1108},{Lowe1112},{Ortin1204},{Kim1205},{Kim1210}}.

Through an interesting observation on the separability of the wave equations, the authors of Refs. \citen{Cvetic1106} and \citen{Cvetic1112} have uncovered a possible geometric realization of the hidden conformal symmetry. This subtracted geometry involves a change of the warp factor in the metric. Although the asymptotic structure is modified, the thermodynamics of the black hole is still preserved (a previous study in Ref. \citen{Yazadjiev0511} may be useful on this point). As explicitly shown in Ref. \citen{Cvetic1201}, the subtracted geometry can also be obtained through a Harrison transformation on the original metric. Recent discussions of the related topics can be found in Refs. \citen{{Virmani1203},{Baggio1210},{Chakraborty1212 1},{Chakraborty1212 2},{Malek1301}}.

Inspired by the above discoveries, one may attempt to find a constructive method which could realize the hidden conformal symmetry through the Harrison metrics. The resulting metrics may also be interesting in their own right. As an elementary 
investigation along this direction, we think it would be valuable and reasonable to focus on the Schwarzschild black hole at first.

The layout of this paper is as follows. In the next section, the relevant results of the Harrison metric proposed in Ref. \citen{Bertini1106} are reviewed. In Sec. \ref{sec 3}, based on a correspondence between the expansion of the Killing vector and the hidden conformal symmetry generators, we give an alternative procedure to obtain this metric. In Sec.  \ref{sec 4}, the subtracted geometry of the Kerr black hole is invoked to find a new Harrison metric for the Schwarzschild case which still preserves the black hole thermodynamics. In Sec. \ref{sec 5}, we discuss the conformal generators for this new metric through the Killing (or Killing's) equations in the near-horizon limit. The conclusion 
can be found in Sec. \ref{sec 6}.

\section{The Harrison metric from hidden conformal symmetry} \label{sec 2}

In this section, we review the hidden conformal symmetry generators for Schwarzschild black hole and the Harrison metric obtained in Ref. \citen{Bertini1106}.

The Schwarzschild black hole is given by the metric
\bea
ds^2 &=& - \f{\D}{r^2} dt^2 + \f{r^2}{\D} dr^2 + r^2(d\th^2 + \sin^2\th d\phi^2),  \label{sch} \\
\D &=& r ( r - r_+ ),
\eea
where the horizon is at $ r_+ = 2M $.
One can easily find the Laplacian as follows
\bea
\ol \na^2 &=& \f{1}{\sr{-g}}\p_{\m}(\sr{-g}g^{\m\n}\p_{\n})  \nn  \\
&=& \f{1}{r^2} [ \ \na^2 + \f{1}{\sin \th} \p_\th \sin \th \p_\th + \f{1}{\sin^2\th} \p^2 _\phi \ ]  ,  \\
\na^2  &=&  \p_r \D \p_r - \f{r^4}{\D} \p^2 _t.
\eea

Following Ref. \citen{Castro1004}, to investigate the hidden conformal symmetry we have to take the near-region, low frequency limit $r \o \ll 1$, $r_+ \o \ll 1$. Here $\o$ is the frequency in the wave function ansatz $\Phi(t,r,\th,\phi) = e^{-i\o t}R(r)Y^l_m(\th,\phi) $. Roughly speaking, all we need here is the correspondence $\p_t \ra - i \o$.

As shown in Ref. \citen{Bertini1106}, in the above limit one can regard the resulting radial Laplacian $\na^2$ as $SL(2,\mathbf R)$ quadratic Casimir. Explicitly, we have
\bea
{\cal H}^2 &=& \f{1}{2}(H_1H_{-1} + H_{-1}H_1) -H_0^2 \nn \\
&=& \p_r \D \p_r - \f{r^4 _+}{\D} \p^2_t .
\eea
Note that as in the literature the radial Laplacian here refers to $\na^2$ while the warp factor part $\f 1 {r^2}$ has been discarded. This is essentially because the latter plays no role in the radial wave equation
\be
\na^2 R =  l ( l + 1 ) R .
\ee

Now we can define the following vector fields as hidden conformal symmetry generators
\bea
H_1 &=& i e^{\f{t}{4M}}\lt( \sr\D \p_r - \f{4M(r-M)}{\sr \D} \p_t \rt)\ , \nn \\
H_0 &=& -i 4M\p_t\ , \label{vec} \\
H_{-1} &=& - i e^{-\frac t{4M}}\lt( \sr\D \p_r + \f{4M(r-M)}{\sr \D} \p_t \rt)\ , \nn
\eea
which satisfy the $SL(2,\mathbf R)$ commutation relations
\be
[H_{\pm 1}, H_0] = \pm i H_{\pm 1}\ , \qquad [H_1,H_{-1}] = 2iH_0 .   \label{com}
\ee

If one naively uses the general formulas in Ref. \citen{Chen1006} and inserts the following parameters for Schwarzschild black hole
\bea
T_L &=& \f{1}{8\pi M} ,  \qquad   T_R = 0  ,  \\
n_L &=& 0 ,  \qquad   n_R = - \f 1{8 M}  ,  \\
\d_+ &=& r - 2 M , \qquad  \d_- = r  ,  \\
A &=& n_L T_R - n_R T_L = \f 1{64 \pi M^2} ,
\eea
then one would also obtain the other set of generators
\bea
\ol H_1 &=& i \lt( \sr \D \p_r  + \f {4M^2}{\sr \D} \p_t  \rt)  ,  \nn  \\
\ol H_0 &=& 0  ,  \\
\ol H_{-1} &=& - i \lt( \sr \D \p_r  - \f {4M^2}{\sr \D} \p_t  \rt)  . \nn
\eea
They obviously do not obey the analogous commutation relations as in Eq. (\ref{com}), and the radial Laplacian could not be reproduced. In this paper, we will follow the interpretation that the Schwarzschild black hole should be described by a chiral conformal field theory (see Ref. \citen{Bertini1106}).

By applying some Harrison transformations on the Schwarzschild metric, the authors of Ref. \citen{Bertini1106} found a new metric. It has a similar structure to the $AdS_2 \times S^2$ Bertotti-Robinson spacetime, and the explicit form is
\bea
ds^2 &=& - e^{-2x}\f{r(r-2M)}{M^2} dt^2 +  e^{2x} \f{M^2}{r(r-2M)} dr^2 +   e^{2x} M^2 (d\th^2 + \sin^2\th d\phi^2). \label{harri 1}
\eea
In order to retain the entropy of the original Schwarzschild black hole, one must set $e^{x} =2$. As pointed out there, the resulting Killing vectors of the $AdS_2$ factor exactly reproduce the hidden conformal symmetry generators in Eq. (\ref{vec}).

\section{Harrison metric and Killing vector}   \label{sec 3}

With the intention to generalize the observation in Ref. \citen{Bertini1106} to other black holes, we propose an alternative procedure which permits us to determine the Harrison metric(s) from the hidden conformal symmetry generators.

Let us start with the following set of $SL(2, \mathbf R)$ generators
\bea
H_1 &=& i e^{\a t + \b \phi} (A \p_r + B \p_t + C \p_\phi) ,  \nn   \\
H_0 &=& - \f {i}{\a} \p_t  ,  \label{gen}  \\
H_{-1} &=& i e^{- \a t - \b \phi} ( - A \p_r + B \p_t + C \p_\phi) . \nn
\eea
Here $\a, \b$ are constants, $A , B , C$ are functions of the parameter $r$, and they all depend on the black hole charges.
Note that for general black holes one would have another set of generators \{$\ol H_1, \ol H_0,\ol H_{-1}$\}, and $H_0$ may acquire an extra term $D \p_\phi$. However, the calculation is quite similar as below.

We assume that the expansion of the Killing vector has a correspondence with the conformal generators as follows
\bea
i \xi  &=&  i  (\xi^r \p_r + \xi^t \p_t + \xi^\phi \p_\phi )  \nn   \\
&=&  a H_1 +  b H_{-1} + c H_0 ,  \label{corre}
\eea
where $a, b ,c$ are some irrelevant constants. Then we can obtain
\bea
\xi^r  &=& A ( a M - b N)  ,  \\
\xi^t  &=& B ( a M + b N) - \f{c}{\a}  ,  \\
\xi^\phi  &=& C ( a M + b N) .
\eea
Here $M \equiv e^{\a t + \b \phi} ,  N \equiv e^{- \a t - \b \phi}$.

Using the Killing (or Killing's) equation
\be
\xi^\r \p_\r g_{\m\n}  + \p_\m \xi^\r g_{\r\n} + \p_\n \xi^\r g_{\r\m}  = 0 ,  \label{kil}
\ee
and treating the metric components \{$g_{tt}, g_{rr}, g_{\phi\phi}, g_{t\phi}$\} as unknown variables,
we arrive at the following equation set
\bea
A \p_r g_{tt} + 2 \a B g_{tt} + 2 \a C g_{t \phi}  &=& 0 ,  \\
A \p_r g_{rr} +  2 \p_r A g_{rr} &=& 0 , \\
A \p_r g_{\phi\phi} +  2 \b C g_{\phi\phi} + 2 \b B g_{t\phi}  &=& 0 ,  \\
A \p_r g_{t\phi}  + (\a B + \b C) g_{t\phi} +  \b B g_{tt} + \a C g_{\phi\phi} &=& 0 ,  \\
\a A g_{rr} + \p_r B g_{tt} + \p_r C g_{t\phi}   &=& 0 ,  \\
\b A g_{rr} +  \p_r C g_{\phi\phi} + \p_r B g_{t\phi} &=& 0  .
\eea
Thanks to the beautiful structure of Eq. (\ref{gen}), each equation here involves a common factor with $a, b$ which has been discarded. If one uses the conformal Killing equation instead, these unwanted constants will appear on both sides of the equations.

Since the first partial derivatives of the metric components are involved,
we would have four integration constants, two of which can be fixed by the last two equations.
This means that we will obtain a family of Harrison metrics for any specific black hole.  By requiring the angular part of the metric structure to be preserved, one can read off the component $g_{\th\th}$. Furthermore, the thermodynamic constraints will determine some integration constants. At the end, the resulting Harrison metric may still not be unique.

Using the hidden conformal symmetry generators of the Schwarzschild black hole given in Eq. (\ref{vec}), the Killing equations are obtained as
\bea
\sr \D \p_r g_{tt} - \f{2(r-M)}{\sr\D} g_{tt} &=& 0 ,  \\
\sr \D \p_r g_{rr} +  2 \p_r \sr \D g_{rr} &=& 0 , \\
\p_r g_{\phi\phi} &=& 0 , \\
\f{\sr \D}{4M} g_{rr} - 4M \p_r \f{r-M}{\sr\D} g_{tt} &=& 0 .   \label{sch ce}
\eea
The solutions can be easily found to be
\be
g_{rr} = \f {K_1}{\D}, \quad g_{tt} = - \f {K_1}{16 M^4} \D , \quad g_{\phi\phi} = K_2 .
\ee
Note that one of the three integration constants has been fixed by the constraint equation (\ref{sch ce}).

To preserve the angular part of the metric structure, we obtain a constraint as $g_{\phi\phi} = g_{\th\th} \sin^2 \th$. On the other hand, for the temperature and entropy to be retained: $T_H = \f 1 {8 \pi M}, S = \f {Area}{4} = 4 \pi M^2$, we must have
\be
K_1 = 4 M^2 , \quad K_2 = 4 M^2 \sin^2 \th .
\ee
In this special case, the requirement about the entropy is equivalent to the condition: $\sr {-g} = \sr {-g_0} \arrowvert_{r_+} $. Here $g_0$ is the determinant of the original metric in Eq. (\ref{sch}).

Thus we have reproduced the Harrison metric in Eq. (\ref{harri 1}) with $e^{x} =2$. More explicitly, this unique metric can be written as
\bea
ds^2 &=& - \f{r(r-2M)}{4 M^2} dt^2 +  \f{4 M^2}{r(r-2M)} dr^2 +  4 M^2 (d\th^2 + \sin^2\th d\phi^2) .
\eea

\section{A new Harrison metric from subtracted geometry}    \label{sec 4}

Although the subtracted geometry of general asymptotically flat black holes in four dimensions has been studied in Ref. \citen{Cvetic1112}, here we only need to extract the corresponding results for the Kerr black hole.

Using the conventions similar to those in Ref. \citen{Cvetic1112}, the metric of the Kerr black hole can be written as follows
\bea
d s^2 &=&  -  \f 1{\sr {\S_0}} G (d t + \f {2 Mr}{G} a \sin^2 \th d \phi)^2 + \sr {\S_0} ( \f{d r^2}{X} + d \th^2 + \f {X}{G} \sin^2 \th d \phi^2)  ,  \label{kerr} \\
X &=& (r-r_+)(r-r_-) = r^2 - 2Mr + a^2  ,   \\
G &=& X - a^2 \sin^2 \th  ,  \\
\S_0 &=& (r^2 + a^2 \cos^2 \th)^2  .
\eea
From this, one can derive the Laplacian
\bea
\ol \na^2 &=& \f{1}{\sr \S_0} [ \ \na^2 + \f{1}{\sin \th} \p_\th \sin \th \p_\th + \f{1}{\sin^2\th} \p^2 _\phi \ ]  ,  \\
\na^2  &=&  \p_r X \p_r - \f{(2Mr \p_t + a \p_\phi)^2}{X} + \f {(2Mr)^2 - \S_0}{G} \p^2 _t.
\eea

The (minimally) subtracted geometry corresponds to the following change of the warp factor
\be
\S_0 \ra \S  =  4M^2 (2Mr - a^2 \cos^2 \th) .
\ee
Now one has the radial Laplacian given by
\be
\na^2  =  \p_r X \p_r - \f{(2Mr \p_t + a \p_\phi)^2}{X} + 4M^2 \p^2 _t ,
\ee
which is exactly the same as the original work on hidden conformal symmetry \cite{Castro1004}.

If one sets the angular momentum $a$ to zero, the Kerr metric (\ref{kerr}) naturally reduces to that of the Schwarzschild black hole in Eq. (\ref{sch}). When the warp factor is changed as $\S_0 \ra \S = 8 M^3 r$, we obtain a new Harrison metric as follows
\bea
ds^2 &=& - \f{r(r-2M)}{2M \sr {2Mr}} dt^2 +  \f{2M \sr {2Mr}}{r(r-2M)} dr^2 +  2M \sr {2Mr} (d\th^2 + \sin^2\th d\phi^2)  .  \label{harri 2}
\eea
The corresponding radial Laplacian is
\be
\na^2  =  \p_r \D \p_r - \f{8 M^3 r}{\D} \p_t ^2 ,
\ee
where $\D = r(r-2M)$.

One can of course define $\r \equiv \sr {2Mr}$, and write the metric (\ref{harri 2}) in another form
\bea
ds^2 &=& - \f{\r(\r^2 - 4 M^2)}{8 M^3} dt^2 +  \f{8 M \r}{\r^2 - 4 M^2} d\r^2
+  2M \r (d\th^2 + \sin^2\th d\phi^2) .   \label{harri 2 2}
\eea
The temperature and entropy can be checked to be the same as the original Schwarzschild black hole.

\section{Conformal generators from Killing equations}    \label{sec 5}

In this section, we derive the (pseudo-)conformal generators for the new Harrison metric given in Eq. (\ref{harri 2}). This heavily relies on the relevant Killing equations in the near-horizon limit.

The following procedure can be regarded as an updated or simplified version of that in Ref. \citen{Franzin1107}. However, some comments on the method given there are in order. Firstly, one can apply the Killing equation directly rather than the conformal Killing equation. Secondly, it is more convenient to use the Killing vectors $\xi^r, \xi^t$ rather than $\xi_r = g_{rr} \xi^r, \xi_t = g_{tt} \xi^t$. Also in this way one does not need to invoke the specific rescaling at the end of that paper.

We firstly use the Killing equation in Eq. (\ref{kil}) to obtain the following equations
\bea
\xi^r \p_r g_{tt} + 2 \p_t \xi^t g_{tt} &=& 0  ,   \\
\xi^r \p_r g_{rr} + 2 \p_r \xi^r g_{rr} &=& 0  ,   \\
\p_t \xi^r g_{rr} + \p_r \xi^t g_{tt} &=& 0  .
\eea
Note that the angular part of the metric is irrelevant as in Ref. \citen{Franzin1107}.
Inserting the parameters given in the metric (\ref{harri 2}), we find the explicit expressions as
\bea
4 \D \p_t \xi^t + (3r - 2M) \xi^r &=& 0  ,  \label{h2 k1}  \\
4 \D \p_r \xi^r - (3r - 2M) \xi^r &=& 0  ,  \label{h2 k2} \\
\D^2 \p_r \xi^t - 8 M^3 r \p_t \xi^r &=& 0  . \label{h2 k3}
\eea

The integration of Eq. (\ref{h2 k2}) leads to
\be
\xi^r = A(t) \ r^{- \f 1{4}} \sr {\D} .  \label{ex r}
\ee
By inserting it in Eqs. (\ref{h2 k1}) and (\ref{h2 k3}), we have
\bea
\p_t \xi^t &=& - A \ \f {3r - 2M}{4 r^{\f 1 {4}} \sr {\D}}  ,  \label{pa t} \\
\p_r \xi^t &=&  \p_t A \ \f {8 M^3 r^{\f 3{4}}}{\D^{\f 3{2}}}  .   \label{pa r}
\eea
With these two equations combined together and noticing the fact $\p_r \p_t \xi^t = \p_t \p_r \xi^t$, we arrive at
\bea
\p^2_t A &-& \wt \l A =  0  ,   \label{eq A}  \\
\wt \l &=& \f {3 r^2 - 4M r + 12 M^2} {128 M^3 r}  .     \label{wt l}
\eea

From Eq. (\ref{pa t}), we can write
\be
\xi^t = - \f {3r - 2M}{4 r^{\f 1 {4}} \sr {\D}} \int^t d t' A (t') + B(r)  .  \label{ex t}
\ee
By inspection of Eq. (\ref{eq A}), we have $\int^t d t' A (t') = \f {\p_t A}{\wt \l}$. So taking the partial derivative of the above equation leads to
\bea
\p_r \xi^t &=& \f {3 r^2 - 4M r + 12 M^2}{16 r^{\f 1{4}} \D^{\f 3{2}} }   \f {\p_t A}{\wt \l}   + \p_r B (r) .
\eea
Recall Eqs. (\ref{pa r}) and (\ref{wt l}), then we observe that $\p_r B (r) = 0$. This means
\be
B (r) =  K',     \label{ex K'}
\ee
which is a constant.

Following Ref. \citen{Franzin1107}, let us take the near-horizon limit $\l \equiv \wt \l \arrowvert _{r_+}$. Then the solution of Eq. (\ref{eq A}) can be found to be
\be
A(t) = \a e^{\sr \l t} + \b e^{- \sr \l t},   \label{ex A}
\ee
where $\a, \b$ are integration constants, and $\sr \l = \f 1 {4M}$ which is equal to the surface gravity of the Schwarzschild black hole.
From the expressions in Eqs. (\ref{ex r}), (\ref{ex t}), (\ref{ex K'}) and (\ref{ex A}), we obtain the Killing vectors as follows
\bea
\xi^r  &=&  r^{- \f 1{4}} \sr {\D} \ (\a e^{\sr \l t} + \b e^{- \sr \l t})   ,   \\
\xi^t  &=&  - \f {3r - 2M}{4 r^{\f 1 {4}} \sr {\D}} \f 1 {\sr \l}  ( \a e^{\sr \l t} - \b e^{- \sr \l t} ) + K'  .
\eea

Using a correspondence similar to Eq. (\ref{corre}), 
we finally arrive at the following pseudo-conformal generators for the new Harrison metric
\bea
H_1 &=& i e^{\f t{4M}} r^{- \f1{4}} \lt( \sr \D \p_r - \f {M(3r - 2M)}{\sr \D} \p_t \rt)  ,   \nn    \\
H_0 &=& K \p_t  =  - i 4M \p_t ,     \\
H_{-1} &=& - i e^{- \f t{4M}} r^{ \f1{4}}  \lt( \sr \D \p_r + \f {M(3r - 2M)}{\sr \D} \p_t \rt) .   \nn
\eea
The previous parameter $K'$ is related to $K = - \f i {\sr \l}$ by another irrelevant constant.
If we define $\wt \D = \sr r (r-2M)$, they can be rewritten as
\bea
H_1 &=& i e^{\f t{4M}}  \lt( \sr {\wt \D} \p_r - \f {M(3r - 2M)}{\sr {r \wt \D} } \p_t \rt)  ,   \nn    \\
H_0 &=& - i 4M \p_t    ,     \\
H_{-1} &=& - i e^{- \f t{4M}}  \lt( \sr {\wt \D} \p_r + \f {M(3r - 2M)}{\sr {r \wt \D} } \p_t \rt)  .   \nn
\eea
These generators clearly have some resemblance with those in Eq. (\ref{vec}).

\section{Conclusion}     \label{sec 6}

The authors of Ref. \citen{Bertini1106} made an interesting observation that the Killing vectors of a specific Harrison metric can reproduce the hidden conformal symmetry generators of the Schwarzschild black hole. As shown in Sec. \ref{sec 3}, by proposing a simple correspondence between the expansion of the Killing vector and the generators, we can treat the metric components as unknown variables and use the Killing equations to find a family of Harrison metrics.

Since the subtracted geometry changes the warp factor in the original black hole and the resulting metric can be obtained through a Harrison transformation, one can find a new Harrison metric for the Schwarzschild black hole using the result of Ref. \citen{Cvetic1112}. The explicit form has been given in Sec. \ref{sec 4}. After that we have derived the pseudo-conformal generators for this new metric by applying the Killing equations in the near-horizon limit. This procedure used in Sec. \ref{sec 5} can be seen as a simplified version of the method in Ref. \citen{Franzin1107}.

It would be interesting to find a refined version of the technique in Sec. \ref{sec 3} which can be generalized to other black holes. This may also give a unified way to describe the minimally and non-minimally subtracted geometries. Note that in Ref. \citen{Cvetic1112}, the subtracted geometry of general asymptotically flat black holes in four dimensions has been connected to the $AdS_3$ space with the metric component $g_{\th\th}$ unaccounted for. Our method also has this disadvantage. However, it may not be an issue since one can still get reasonable results by resorting to the thermodynamic requirements. Also this is actually related to the general feature of the hidden conformal symmetry generators.

The new Harrison metric in Eq. (\ref{harri 2}) (or Eq. (\ref{harri 2 2})) may provide a bridge between the Schwarzschild black hole and the Kerr black hole. The previous work in Ref. \citen{Clement1997} may be relevant for this idea. On the other hand, the physical implications of this metric deserve further investigation. For example, one can use the traditional methods in Ref. \citen{Berti0905} to find the quasinormal modes. However, to resolve the problem pointed out in Ref. \citen{Kim1210} within the hidden conformal symmetry approach, one may need to improve the operator method proposed in Ref. \citen{Chen1009}. Last but not the least, one may follow the works in Refs. \citen{{Baggio1210},{Chakraborty1212 1},{Chakraborty1212 2},{Malek1301}} to find more connections between the subtracted geometry approach and other areas.

\section*{Acknowledgments}

This work was supported by National Natural Science Foundation of
China (No. 11173028 and No. 10875009).

\appendix

\end{document}